\documentstyle[12pt]{article}
\let\section=\subsection     \let\subsection=\subsubsection                
\newcommand\noi{\noindent}
 \newcommand\beq{\begin{equation}}
 \newcommand\eeq{\end{equation}}
 \newcommand\beqn{\begin{eqnarray}}
 \newcommand\eeqn{\end{eqnarray}}
\newcommand{\doublespace} {
 \renewcommand{\baselinestretch} {1.6}
 \large\normalsize}

\begin{document}

\vspace*{4cm}
\begin{center}
   {\large \bf DYNAMICS AND PHENOMENOLOGY}\\[2mm]
   {\large {\bf OF CHARMONIUM PRODUCTION OFF NUCLEI}
\footnote{
Based on talks presented at the Workshop on Heavy Ion Collisions,
Clermont Ferrand, France, January 9-10, 1997 and at the Workshop 
'QCD Phase Transitions', Hirschegg, 
Austria, January 13-18, 1997}}\\
[5mm]
BORIS Z. KOPELIOVICH \\[5mm]
   {\small \it  Max-Planck-Institut f\"ur Kenphysik\\
   Postfach 103980, D-69029 Heidelberg, Germany \\ }
	{\small \it	and}\\
{\small \it Joint Institute for Nuclear Research\\
141980 Dubna, Moscow Region, Russia\\[8mm]}
\end{center}

\vspace*{2cm}

\begin{abstract}\noindent
Nuclear suppression of charmonium production in proton-nucleus
interactions is poorly understood, what restrains
our attempts to single out unusual effects in 
heavy ion collisions. We develop a phenomenological approach,
based on the light-cone dynamics of charmonium
production, which has much in common with 
deep-inelastic scattering and Drell-Yan lepton 
pair production. The key observation is the existence of a
 soft mechanism of heavy flavour
production, which scales in the quark mass and dominates
shadowing corrections and diffraction.
It naturally explains the surprisingly strong nuclear suppression
of $J/\Psi$ at large Feynman-$x_F$. The low-$x_F$ region is subject 
to a complicated interplay of hard and soft mechanisms.
With evaluated parameters we nicely describe available data
on charmonium production in proton-nucleus collisions.
Using these results we predict a new process, diffractive
production of charmonium on a nucleon target, which
fraction in the total production rate of charmonium is 
evaluated at $12\ \%$.
   
\end{abstract}

\doublespace
\newpage

\section{Introduction}
Production of charmonium in heavy ion collisions is believed
to be a sensitive probe for new physics. 
In practice, however, hand-waving 
motivations are not sufficient, and one desperately needs a
reliable base line. 
Unfortunately, present understanding of dynamics of hadroproduction
of charmonium off protons and nuclei
is far from satisfactory.
Nuclear effects are poorly understood, particularly,
there is still no reasonable quantitative explanation of
surprisingly strong nuclear suppression of $J/\Psi$ production at high
Feynman $x_F$. It is too adventurous to predict or to interpret the
observed nuclear effects in heavy-ion collisions, having no idea
about what is going on in proton-nucleus collisions.

I do not pretend in this short talk to 
establish a reliable baseline, but
only want to make another step towards it. 

This talk is based mostly on the 
recent unpublished results obtained in
collaboration with J\"org H\"ufner, as well as on
our previous publications, cited in the text.

\section{Poor man's formula or how long does it 
take to produce a $c\bar c$ pair}

There is a wide spread believe, 
adopted in most recent analysis of data
on $J/\Psi$ production, that the $c\bar c$ pair 
is produced momentarily at the point of interaction of the
projectile hadron with a bound nucleon, and then it propagates
through the nucleus and eventually escapes it. 
This point of view is inspired by the usual perturbative 
treatment of $c\bar c$
production in the rest frame 
of the charmonium. However, in the rest
frame of the nucleus the production time is subject to Lorentz time dilation
(or the nuclear thickness is contracted 
in the charmonium rest frame). When the production time is
longer than the mean internucleon distance in nuclei ($\sim 2\ fm$)
one cannot say anymore that the $c\bar c$ pair is produced on
one concrete nucleon. This can also be understood as coherence
between the waves produced at different nucleons. For this reason 
the production time is usually called coherence time (length).

The time scale of charmonium production has two extremes.
In the low energy limit the coherent time $t_c \ll d$, where
$d$ is the mean internucleon distance,one can use 
a simple (usually called a conventional approach) formula for
nuclear suppression,
\beqn
S^{\Psi}_{hA} &=& {1\over A}\ 
\int d^2b \int\limits_0^{\infty} dz\ \rho(b,z)\ 
\exp[- \sigma_{in}^{\Psi N}T_z(b)] =
\frac{\sigma_{in}^{\Psi A}}{\sigma_{in}^{\Psi N}}
\nonumber\\
&\approx& 1-{1\over 2}\sigma_{in}^{\Psi N}\ 
\langle T\rangle\ .
\label{1}
\eeqn
\noi
Here the mean nuclear thickness 
$\langle T\rangle = A^{-1}\int d^2b\ T^2(b)$, where
$T(b)= T_z(b)|_{z\to -\infty}$ and
$T_z(b) = \int_z^{\infty}dz\ \rho(b,z)$.
The nuclear density $\rho(b,z)$ depends on impact parameter 
$b$ and longitudinal coordinate $z$. The latter approximation 
in (\ref{1}) uses smallness
of the charmonium inelastic interaction cross section,
$\sigma_{in}^{\Psi N}\ T(b) \ll 1$.

Although a $c\bar c$ pair, rather than the $\Psi$, is
produced and propagates through the nucleus, in the low energy
limit one may think that the charmonium wave function is formed
instantaneously. We come later back to this point and formulate an 
exact condition for that.

This simple formula corresponding to the low-energy limit is
nowadays widely used as a phenomenological basis for
nuclear effects in charmonium production. The absorption cross section
$\sigma_{in}^{\Psi N}$ is usually treated as an unknown parameter.

In the high energy limit, $t_c \gg R_A$, a fluctuation,
containing the $c\bar c$ pair lives much longer than the 
nuclear size. The nuclear suppression factor has in this case a 
form similar to that for quasielastic scattering \cite{kz91},
\beq
S^{\Psi}_{hA}|_{l_c\gg R_A}= 
\frac{1}{A\ \sigma_{free}}\ 
\int d^2b\ T(b)\ \exp[-\sigma_{free}\ T(b)]
\approx 1-\sigma_{free}\ \langle T\rangle\ .
\label{2}
\eeq
\noi
I would like to draw attention to the fact that
in the approximation $\sigma_{in}^{\Psi N}\ T(b) \ll 1$ 
the shadowing term in (\ref{1}) has a factor $1/2$ 
compared to that in (\ref{2}). This is an explicit
manifestation of the space-time pattern of interaction: at high
energy the mean length of path of the $c\bar c$ pair in the nucleus is
twice as long as at low energy.

Let us postpone to the next section
interpretation of the freeing cross section $\sigma_{free}$,
but for now we
concentrate on the problem of the coherence length.
The lifetime of a fluctuation of the projectile hadron
containing the $c\bar c$ pair is given by the 
energy denominator,
\beq
t_c = \frac{2E_h}{M^2_{fl} - m_h^2}\ ,
\label{3}
\eeq
\noi
where $E_h$ and $m_h$ are the energy and
mass of the projectile hadron, and $M_{fl}$ is
the effective mass of the fluctuation, which is
to be considered on mass shell in the light-cone
formalism. The kinematical formula
for $M_{fl}^2$ reads,
\beq
M_{fl}^2 = \sum\limits_{i} 
\frac{m_i^2 + k_i^2}{\alpha_i}\ ,
\label{4}
\eeq
\noi
where $m_i$, $k_i$ and $\alpha_i$ are the mass, 
the transverse momentum and the fraction of 
the total light-cone momentum carried by the 
$i$-th parton, and we sum over all the partons
in the fluctuation ($\sum_i \alpha_i = 1$).

It is easy to find from (\ref{4}) 
the minimal effective mass
of a fluctuation containing a $c\bar c$ pair, which
corresponds to a longest coherence time,
\beq
t_c \leq \frac{2E_{\Psi}}{M^2_{\Psi}}
\label{5}
\eeq
\noi
This is the most general upper restriction for the lifetime of
any fluctuation containing a $c\bar c$ pair. 

It turns out that at energies corresponding to available data
on $\Psi$ production we have a full coverage of possible
scenarios of space-time development. 
For instance at 
$x_F \approx 0$ we have $t_c \leq \sqrt{s}/2m_NM_{\Psi}$.
This is only about $1\ fm$ at $E_h = 200\ GeV$, however, $t_c$
becomes 
compatible with the radiuses of heavy nuclei at Tevatron energies,
$E_h \approx 800\ GeV$, and it is much longer in the latter case if
$x_F \to 1$. 

It is demonstrated on different exact solutions in 
\cite{bkmnz,hkn,hkz} that in the approximation
$\sigma_{in}^{\Psi N}\langle T\rangle \ll 1$ 
the variation of nuclear
suppression as function of the coherence length 
always has a form of a linear function of the
nuclear longitudinal formfactor squared,
\beq
F^2_A(q_c) = \frac{1}{\langle T\rangle}\ 
\int d^2b\ \left|\int\limits_{-\infty}^{\infty}
dz\ \rho(b,z)\ e^{iq_cz}\right|^2\ ,
\label{5a}
\eeq
\noi
where the longitudinal momentum transfer $q_c = 1/t_c$.
This formfactor varies from zero in the low-energy
limit $t_c \ll R_A$ up to one at $t_c \gg R_A$.
The formula for nuclear suppression, which interpolates
between the low- and high-energy limits, (\ref{1}) and 
(\ref{2}), reads,
\beq
S^{\Psi}_{hA} = 1 -
{1\over 2}\sigma_{in}^{\Psi N}\
\langle T\rangle\left[1-F^2_A(q_c)\right]
- \sigma_{free}\ \langle T\rangle\ 
F^2_A(q_c)
\label{5b}
\eeq
In the two following sections we elaborate
more with the cross sections $\sigma_{free}$ and 
$\sigma_{in}^{\Psi N}$.

\section{The freeing cross section and the soft 
mechanism of charmonium production}

The soft spectator mechanism of heavy 
flavour production was the subject
of my talk in Hirschegg in 1995 \cite{hir95}.
Let me summarize the observations.

First of all, one should discriminate 
between the interaction cross
section of a hadronic  fluctuation, 
containing a heavy quark pair and the
cross section of freeing (production of) this pair. 
The former is large, since the fluctuation as a whole is
big, while the latter
is usually small, of the order of the 
inverse heavy quark mass squared. Indeed,
soft (low $k_T$) long-wave (in transverse direction) gluons, 
providing a large cross 
section, cannot resolve such a small-size structure
as a heavy-quark fluctuation. The freeing amplitude can be
represented as a difference between interaction amplitudes
of two Fock states, which are identical. except one 
contains the $c\bar c$ fluctuation, 
but another one does not. This is the
general principle originating from the pioneering ideas
of Feinberg-Pomeranchuk and Good-Walker 
for diffractive production. If 
all the fluctuations of the projectile hadron have
the same interaction amplitudes, nothing new
can be produced, but the initial hadron, which coherence
is not disturbed in this case. 

Let us consider a concrete example of a fluctuation of a light quark,
which consists of the same quark and a colorless $c\bar c$ pair.
The freeing amplitude is equal to the difference between the amplitude
of inelastic interaction of this fluctuation and that of the single
quark. Although each of these amplitudes is infrared divergent,
their difference is finite. Since only the light quark-spectator
can interact (interaction of the colorless $c\bar c$ pair would lead 
either to open flavour production, or is a higher order 
$1/m_Q^2$ effect) the difference of the inelastic amplitudes 
comes from the difference of the impact parameters of the quark
with and without the $c\bar c$ fluctuation. 
If the transverse separation between the quark and the center of 
mass of the $c\bar c$ pair is $r_T$, the center of gravity of the
whole fluctuation (which coincides with the impact parameter of the
parent light quark) is distant by 
$(1-\alpha)r_T$ from the $c\bar c$ pair and by $\alpha r_T$ from
the light quark, where $\alpha$ is the fraction of the light-cone
momentum of the projectile quark. 
Thus, the {\it freeing} cross section equals to
the total {\it interaction} cross section of a light $q\bar q$ 
pair with separation $\alpha r_T$. Therefore, at small $r_T$ 
color transparency leads to $\sigma_{free}(r_T) \propto r_T^2$.
This was an important observation
of \cite{hir95} ( see also \cite{bhq}).

The mean transverse separation $\langle r_T\rangle$ 
between the $c\bar c$ and the $q$ can be estimated as follows 
\cite{hir95}. The wave function of the fluctuation 
in the momentum representation is given by
the energy denominator, $\Psi_{c\bar cq}(k_T) \propto
1/(M_{fl}^2-m_q^2) \propto 1/(k_T^2 + M_{c\bar c}^2(1-\alpha)
+ \alpha^2 m_q^2)$, where (\ref{4}) is used for $M_{fl}^2$.
Switching to the impact-parameter representation 
(only for the transverse momentum) one gets the wave 
function of the $qc\bar c$ fluctuation (the spin structure
is neglected) $\Psi_{c\bar cq}(r_T,\alpha) \propto K_0(\tau r_T)$ 
\cite{hir95}, where $\tau^2 = (1-\alpha)
M_{\Psi}^2 + \alpha^2 m_q^2$. This distribution gives the
mean separation $\langle r_T^2\rangle \sim 1/\tau^2$, and
the freeing cross section reads,
\beq
\sigma_{free}(\alpha) \sim \frac{\alpha^2}
{(1-\alpha)M_{\Psi}^2 + \alpha^2m_q^2}\ ,
\label{6}
\eeq
As different from naive expectations, eq.~(\ref{6}) says
that the cross section of freeing of a heavy quark fluctuation
is small unless the factor $(1-\alpha)$ is small 
$\sim m_q^2/M_{\Psi}^2$, i.e. the $c\bar c$ takes the main fraction of
the initial quark momentum. In this case the freeing cross section
is as large as a typical hadronic one, $\sim 1/m_q^2$ (actually,
$m_q$ should be treated as an effective infra-red cut off 
for $r_T$ of the 
order of the inverse confinement radius, rather
than the quark mass).
This fact may have relevance to the well known
problem of a substantial increase of nuclear suppression of
$J/\Psi$ production at large $x_1 = p_{\Psi}^+/p_{h}^+$, which 
corresponds to an effective absorption cross section close to that
for light hadrons. We come back to this problem below.

The production charmonium cross section is given by a 
convolution of the cross section
(\ref{6}) with the momentum distribution function $ F_h(y)$
of the quark in the projectile hadron (more generally,
the distribution of those partons, which participate in 
the fluctuation containing the $c\bar c$).
\beq
\langle\sigma^h_{free}(x_1)\rangle \propto
\int d^2r_T\ \int\limits_{x_1}^{1} 
dy\ \left|\Psi_{c\bar cq}(r_T,x_1/y)\right|^2
\ F_h(y)\ \sigma_{free}(r_T)
\label{7}
\eeq
\noi
Assuming the end-point behaviour for $F_h(y)\propto (1-y)^{\beta}$,
one arrives to the following $x_1$-dependence of the 
freeing cross section at large $x_1$,
\beq
\langle\sigma^h_{free}(x_1)\rangle \propto
\frac{(1-x_1)^{\beta}}{M^2_\Psi}\ 
\left\{\ln\left(\alpha_s
\left[(1-x_1)M_{\Psi}^2/m_q^2\right] 
+ c\right)\right\}\ .
\label{8}
\eeq
\noi
The double-logarithmic dependence on $M_{\Psi}^2$
corresponds to the well known violation of the Bjorken
scaling in deep-inelastic scattering 
given by the evolution equations. However, the hard scale
in (\ref{8}) is imposed by $(1-x_1)M_{\Psi}^2$, rather
than $M_{\Psi}^2$, prescribed by the conventional 
factorization theorem. Another violation of the factorization
is due to the constant $c$ in (\ref{8}), which depends on
the exponent $\beta$ in the hadronic structure function.

Amazingly, the contribution of the soft asymmetric
fluctuations with $1-\alpha \sim m_q^2/M^2_{\Psi}$, corresponding
to large freeing cross section (see (\ref{7})), does not
vanish at high $M^2_{\Psi}$. Moreover, these soft fluctuations
dominate in nuclear shadowing, which is given by the same expression
(\ref{8}), but with higher powers of $\sigma_{free}$.
This can be explained in a simple way
(like in deep-inelastic scattering 
\cite{kp97})if
to divide all the fluctuations to two classes, the hard one 
with $\sigma_{free} \sim 1/M^2_{\Psi}$, and the soft one 
with $\sigma_{free} \sim 1/m_q^2$. The results for the production
cress section and the first order shadowing correction are summarized
in Table~1.

\vspace{.2cm}
\begin{center}
Table 1.
Contributions of soft and  hard fluctuations
to the charmonium\\  production
cross section
and to nuclear shadowing

\vspace{.3cm}
{\doublespace

\begin{tabular}{|c|l|l|l|l|c|}
 \hline
Fluctuation & $\left|\Psi_{c\bar cq}\right|^2$ 
& $\sigma^{hN}_{tot}$ &
$\left|\Psi_{c\bar cq}\right|^2
\sigma^{hN}_{tot}$ &
 $\left|\Psi_{c\bar cq}\right|^2
(\sigma^{hN}_{tot})^2$\\
 \hline Hard & $\sim 1$ & $\sim 1/M^2_{\Psi}$
& $\sim 1/M^2_{\Psi}$  & $\sim 1/M_{\Psi}^4$\\
 \hline
 Soft & $\sim m_q^2/M^2_{\Psi}$ &$\sim1/ m_q^2$&
$\sim 1/M^2_{\Psi}$ & $\sim 1/ m_q^2M^2_{\Psi}$\\
\hline
\end{tabular}
}
\vspace{.3cm}
\end{center}

One can see from the Table that the soft component dominates
the first-order nuclear shadowing term proportional to 
$\langle[\sigma_{hN}^{\Psi}(r_T)]^2\rangle$,
which 
has the same $ 1/M^2_{\Psi}$ suppression and the 
the impulse approximation term 
$\langle\sigma_{hN}^{\Psi}(r_T)\rangle$.
This means that shadowing scales in $M^2_{\Psi}$.
This is the common feature of deep-inelastic scattering 
\cite{nz91},
Drell-Yan pair and heavy flavour production \cite{hir95},
where the soft and hard contributions are known to be
of the same order.

Although the dynamics of charmonium production in
hadronic collisions is much more complicate,
one may try to develop a phenomenological 
approach relying on the general ideas on space-time development
(section 2) and on the dynamics of nuclear attenuation (this
section).

Let us classify different contributions to nuclear suppression
factor (\ref{5b}).

{\bf 1.}
Since the dominant contribution to
nuclear shadowing comes 
from asymmetric fluctuations $q\to c\bar cq$ 
with $\alpha \to 1$,
the $J/\Psi$ has the same $x_1$-distribution
as the valence quark, i.e. $\sim (1-x_1)^3$ 
if the projectile is a proton. This includes also the
so called color-octet mechanism which corresponds to
the fluctuation $q\to qc\bar c2g$.
The exponent in the $x_1$-distribution may be even smaller if
the colorless $c\bar c$ fluctuation is produces collectively
by two or three valence quarks, as was suggested in \cite{bhmt}.

Note that in the dominant asymmetric fluctuation $1-\alpha
\approx m_q^2/M_{\Psi}^2$ and the light quark contributes
to the effective mass of the fluctuation (\ref{4}) 
the same amount as 
the $c\bar c$ pair. Therefore, the formfactor $F_A^2(q_c)$
suppressing the soft fluctuations in (\ref{5b}) should be
evaluated at $q_c \approx 2q_{\Psi}$, where $q_{\Psi}=
M_{\Psi}^2/2E_{\Psi}$ is according to (\ref{5}) the minimal
longitudinal momentum transfer.

The soft freeing cross section $\sigma_S$ can
be borrowed from the data on nuclear shadowing in 
deep-inelastic scattering.
In accordance with the modified factorization
relation (\ref{8}) we use the result of analysis 
\cite{kp} of the data on deep-inelastic scattering 
at $Q^2=(1-x_1)M_{\Psi}^2$
and fix, $\sigma_S \approx 12\ mb$ ,
assuming $x_1 \approx 0.5$. However, at $x_1>0.9$ 
(where no data are still available) the corresponding value 
of $Q^2$ is so small that on gets into the domain of the
vector dominance model, and $\sigma^S_{free}$ should be about
$20-30\ mb$

{\bf 2.}
Hard mechanisms of charmonium production include 
direct interaction of the color-octet $c\bar c$ fluctuation 
with the target (e.g. the color-singlet mechanism).
Due to color screening the freeing cross section is $\propto 
\langle r^2_{c\bar c}\rangle \approx 4/m_c^2\approx 0.07\ fm^2$
(the factor $4$ is due to color screening suppressing the
fluctuations of small size \cite{kz91}). Estimated perturbatively 
$\sigma_H \approx 2\ mb$, substantially 
smaller than one expects for $\sigma_{in}^{\Psi N}$.

Since the effective mass of the hard fluctuation is $\sim
M_{\Psi}$, the formfactor $F_A^2(q_c)$ in (\ref{5b}) should be 
taken at $q_c \approx q_{\Psi}$

The produced charmonia have the same $\propto (1-x_1)^5$ 
distribution as the parent gluon.
Thus, the hard contribution to the
freeing cross section
is suppressed by $(1-x_1)^n$ compared with the soft mechanism,
where $n\geq 2$.

{\bf 3.}
The inelastic $J/\Psi$-nucleon cross section is known
from experiment (see in \cite{kz91}) with a large uncertainty
and ranges from 2 to 7 $mb$. The lowest-order, two gluon
graph provides $\sigma_{in}^{\Psi N}\approx 5\div 6\ mb$.

We expect quite a steep energy dependence of $\sigma_{in}^{\Psi N}$.
Indeed, $J/\Psi$ has a smaller radius than the light hadrons,
and perturbative QCD predicts a growth of the effective
Pomeron intercept with decreasing radius. The recent measurements
of energy dependence of the cross section of elastic photoproduction
of $J/\Psi$ at HERA \cite{warsaw} found that it grows 
$\propto s^{0.4}$. Since this cross section is 
$\propto \sigma_{el}^{\Psi N}\approx 
[(\sigma_{tot}^{\Psi N}]^2/16\pi B$, 
one can find the energy dependence of $\sigma_{tot}^{\Psi N}
\propto s^{0.25}$, taking into account the energy dependence of the 
slope parameter $B$. Note that such a high effective Pomeron intercept
perfectly agrees with the measured at HERA $x$-dependence of
the proton structure function $F_2^p(x,Q^2)$
at the corresponding virtuality
$Q^2 \approx M_{\Psi}^2$.

The steep energy dependence of $\sigma_{in}^{\Psi N}$
results in a substantial variation of nuclear suppression,
provided by the first term in eq.~(\ref{5b}).

We expect the same energy-dependence for the hard component,
while the soft contribution 
is supposed to grow slowly, $\propto s^{0.1}$.

\section{Evolution of a $c\bar c$ wave packet in a medium.
The $\Psi/\Psi'$ puzzle}

The low-energy limit (\ref{1}) was written assuming that
the charmonium is produced momentarily at the point
of interaction with a bound nucleon. However, a $c\bar c$ 
pair is produced, rather than the charmonium, and it 
takes some time to become the final charmonium, even if
$t_c \ll R_A$. The evolution of the $c\bar c$ wave packet
is controlled by a different parameter, called formation
time/length, 
\beq
t_f \geq \frac{2E_{\Psi}}{M_{\Psi'}^2- 
M_{\Psi}^2}\ ,
\label{9}
\eeq
\noi
which is about five times longer than $l_c$.

The evolution can be solved using either the path integral technique
in quark representation \cite{kz91} or, what is equivalent,
in hadronic representation using the coupled-channel approach 
\cite{hk95}. In the latter case the two channel approximation, 
$J/\Psi$ and $\Psi'$, turns out to have a pretty good, 
about $10\ \%$ accuracy. In this case the effective
cross section of  absorption of $J/\Psi$ in (\ref{1}) and ({\ref{5b})
should be replaced by
\beq
\sigma_{in}^{\Psi N}\ \Longrightarrow\ 
\sigma_{in}^{\Psi N}\ \left[1 + 
\epsilon R\ F_A^2(q_f)\right]\ ,
\label{10}
\eeq
\noi
where $q_f = 1/l_f$, and $F_A^2(q)$ is defined in (\ref{5a}).
Other parameters are defined in \cite{hk95}, $R^2\approx 0.25$ is
the experimentally known 
ratio of $\Psi'$ to $J/\Psi$ production rates. $\epsilon=-\sqrt{2/3}$
is the ratio of the off-diagonal to the diagonal diffractive $\Psi - N$ 
amplitudes evaluated in \cite{hk95}.

With these parameters one concludes that if $l_f\gg R_A$
the effective absorption cross section (\ref{10}) is only about
$0.6$ of that for $J/\Psi$. 

Another interesting observation of \cite{hk95} 
concerns nuclear suppression of 
$\Psi'$. Naively, using the poor man's formula (\ref{1})
one would expect much stronger suppression of the $\Psi'$
than $J/\Psi$. However,
the corresponding effective absorption cross section
turns out to be quite different from $\sigma_{in}^{\Psi' N}$,
\beq
\sigma_{in}^{\Psi' N}\ \Longrightarrow\
\sigma_{in}^{\Psi' N}\ \left[1 +
\frac{\epsilon}{r R}\ F_A^2(q_f)\right]\ ,
\label{11}
\eeq
\noi
where $r =\sigma_{in}^{\Psi' N}/\sigma_{in}^{\Psi N}\approx 7/3$
as evaluated in \cite{hk95}. It is interesting that with these
parameters the effective absorption cross sections (\ref{10})
and (\ref{11}) are approximately equal, provided that
$l_f \gg R_A$, i.e. $F_A^2(q_f)=1$. This observation naturally
explains the surprising experimental result of nearly the same
nuclear suppression of $J/\Psi$ and $\Psi'$.
Note that in the case of the soft mechanism only the light 
projectile quark interacts with the target, and the structure of
the $c\bar c$ pair also does not affect the nuclear attenuation.

\section{Phenomenology of $J/\Psi$ production off nuclei}

Summarizing the results of previous sections, eq.~(\ref{5b})
can be presented in the unitary form,
\beq
S^{\Psi}_{hA} = \exp\left[-\sigma_{eff}(E_h,x_1)\ 
\langle T\rangle\right]\ ,
\label{12}
\eeq
\noi
where
\beqn
\sigma_{eff}(E_h,x_1)& = &
{1\over 2}\sigma_{in}^{\Psi N}
\left[1 +
\epsilon R\ F_A^2(q_{\Psi})\right]
\left[1-F^2_A(q_{\Psi})\right] 
\nonumber\\
& - &\frac{(1-x_1)^n F_A^2(q_{\Psi})\sigma_H^2 +
\gamma F_A^2(2q_{\Psi})\sigma_S^2}
{(1-x_1)^n\sigma_H +
\gamma\sigma_S}\ .
\label{13}
\eeqn
\noi
All the notations here were introduced before, except
$\gamma$, which is the ratio of the weight factors
of the soft to the hard components. According to Table~1
this factor is suppressed by the $J/\Psi$ mass,
$\gamma \sim m_q^2/M^2_{\Psi} \approx 0.004$, if to
treat $m_q$ as a cut off, which is the inverse confinement 
radius $\sim \Lambda_{QCD}$.
Other parameters are also pretty well known.
$\sigma_{in}^{\Psi N}= \sigma_{\Psi}^0\ (x_1 E_h)^{\Delta}$,
where $\Delta = 0.25$ and the parameter $\sigma_{\Psi}^0$ 
should provide the cross section about $5\div 6\ mb$ at the energy
few tens $GeV$. Then we expect $\sigma_{\Psi}^0\sim 2\ mb$.
The hard cross section $\sigma_H = \sigma_H^0\ 
(x_1 E_h)^{\Delta}$, as was mentioned, 
is expected to be smaller than $\sigma_{in}^{\Psi N}$, 
of the order of $2\ mb$. Then we expect $\sigma_H^0\sim 0.7\ mb$. 
The exponent $n\geq 2$ is not very important, 
we compare with the data at $n=3$. 

To elaborate with low-energy data on $J/\Psi$ production 
one should take care of
possible effects of induced energy loss, which are due to
rescattering of the projectile partons in the nucleus \cite{kn}.
This is unimportant at the energy $E_p=800\ GeV$, where the best
data exist, but may produce an additional suppression at lower energies.
We include this effect in a rough way through the extra factor $K(E_h,x_1)$
to the expression (\ref{12})
\beq
K(E_h,x_1) = \left(\frac{1- \widetilde x_1}
{1- x_1}\right)^m\ .
\label{14}
\eeq
\noi
Here $\widetilde x_1 = \Delta E/E_h$ and
$\Delta E = 
\kappa\ \langle L\rangle$, where $\langle L\rangle
\approx 3R_A/4[1+F^2_A(q_{\Psi})]$ is the mean free path
of a parton in the nucleus, where we took into account
the Lorentz stretching of the path of a fluctuation.
The density of energy loss $dE/dz=-\kappa$ was fixed at
the color-triplet string tension $\kappa=1\ GeV/fm$.
The exponent $m=5$ in accordance with the measured
$x_1$ dependence of $J/\Psi$ production rate in $pp$ collisions.

\begin{figure}[tbh]
 \includegraphics{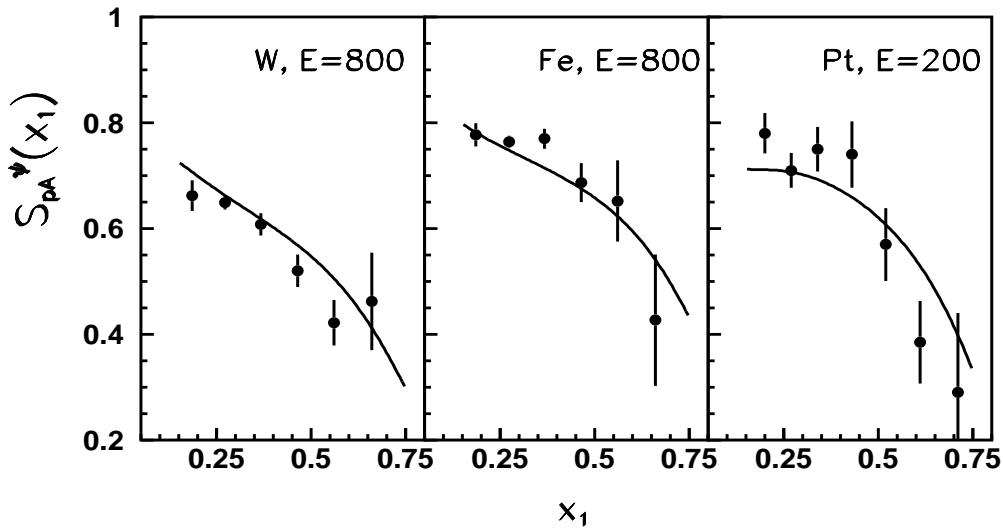}
\begin{center}
\vspace{6.5cm}
\parbox{13cm}
 {\caption[Delta]
 {\it Nuclear suppression of $J/\Psi$
production rate as measured in \cite{e772,na3}
and calculated with 
eq.~(\ref{12}). See explanations in the text.}
\label{fig1}}
\end{center}
\end{figure}

Fitting the data depicted in Fig.~1 
on nuclear suppression from the experiments
E772 at $800\ GeV$ \cite{e772} and
NA3 at $200\ GeV$ \cite{na3} in
$pA$ collisions,
as well as the data 
of the NA38/50 experiments at $200$ ($\bar x_1\approx 0.25$)
\cite{na38} 
and $450\ GeV$ ($\bar x_1\approx 0.1$) \cite{na50}
 we arrived at the 
following best values of the free parameters in (\ref{13}),
$\gamma = 0.0037 \pm 0.0015$,
$\sigma^0_{\Psi}=1.95 \pm 0.018\ mb$ and
$\sigma^0_H=0.64 \pm 0.05\ mb$, which are 
in a perfect agreement with our
pre-evaluations. Comparison with the data, 
in Fig.~1 and Table~2
 demonstrates 
a nice agreement.

\vspace{.2cm}
\begin{center}
Table 2.
Data on $pA$ collisions 
from the NA50 experiment at $450\ GeV$
\cite{na50} and from the NA38 experiment 
at $200\ GeV$ \cite{na38} 
compared to our results (bottom row)
\vspace{.3cm}
{\doublespace

\begin{tabular}{|c|l|l|l|l|l|c|}
 \hline
450, W
& 450, Cu  &
450, Al &
450, C & 200, U & 200, W & 200 Cu \\
 \hline $0.67\pm 0.08$  &$ 0.75\pm 0.08$
&$ 0.76\pm 0.09$  &$ 0.85\pm 0.1$ &$ 0.63\pm 0.13$ &
$0.65\pm 0.05$ &$ 0.75\pm 0.17$ \\
 \hline
0.74 & 0.82 &
0.87 & 0.90 & 0.69 & 0.71 & 0.80 \\
\hline
\end{tabular}
}
\vspace{.3cm}
\end{center}
 
Although our results agree with the 
NA38/50 data within error bars,
they are yet systematically higher than the nominal 
experimental values. This is mostly a result of a possible 
inconsistency between the data from the NA3 and NA38/50
experiments. The former are higher and have quite
small error bars (see Fig.~1). This region
of low $x_1$ is controlled by the values of
$\sigma_{\Psi}^0$ and $\sigma_H^0$, which are the free
parameters and do not need to vary much in order
to fit better either of the data.

\section{Diffractive production of charmonia}

Using the data on nuclear shadowing one can 
estimate the cross section of diffractive production
of charmonium in hadron-nucleon interaction, i.e.
diffractive excitation of the projectile hadron
to a state containing the charmonium and light hadrons,
while the target nucleon is intact.

It is well known since Gribov's paper \cite{gribov}
that nuclear shadowing of the total cross 
section is closely related to 
diffraction. It is a direct consequence of unitarity
that the forward diffractive cross section
can be represented as

\beq
\int dM^2\left.\frac{d\sigma_{DD}}
{dp_T^2 dM^2}\right|_{p_T=0}=
\frac{1}{16\pi}\ 
\langle\sigma^2\rangle
\label{14a}
\eeq
\noi
one averages  here over the eigen states of interaction,
which may be treated as the projectile fluctuations.
The cross section (\ref{14a}) includes also the elastic
channel.

We are interested only in that part of diffraction in
(\ref{14a}), which contains a colorless $c\bar c$ pair
in final state. In this case one should replace the 
interaction cross sections in (\ref{14a}) by the freeing 
cross sections. Then one arrives at the same expression,
which we had for nuclear shadowing term in charmonium 
production (see section 3) when the nuclear formfactor
saturates ($q_cR_A \ll 1$),
\beq
1 - S_{hA}^{\Psi} =
\frac{16\pi}{\sigma(hp\to \Psi X)}\ 
\langle T\rangle\left. 
\int\limits_{(m_N + M_{\Psi})^2}^{M^2_m}
dM^2\ \frac{d\sigma^{\Psi}_{DD}}{dM^2dp_T^2}
\right|_{p_T=0}\ ,
\label{14b}
\eeq
where $d\sigma^{\Psi}_{DD}/dM^2dp_T^2$
is the differential cross section of diffractive production
of charmonium in $hN$ collisions, i.e. the process
$hN\to X_{\Psi}N$, where the multiparticle state $X_{\Psi}$ 
consists of $\Psi$ and light
hadrons. $M$ is the effective mass of the $X_{\Psi}$, which 
is restricted by the nuclear formfactor at $M^2 < M_m^2
\sim E_h/R_A$.

As soon as nuclear shadowing of $\Psi$
production is known, we can predict the fraction of 
diffraction in the 
total cross section of $\Psi$ production,
which reads
\beq
\delta^{\Psi}_{DD}(M_m) \equiv
\frac{1}{\sigma(hp\to \Psi X)}\ 
\int dp_T^2
\int\limits_{(m_N + M_{\Psi})^2}^{M^2_m}
dM^2\ \frac{d\sigma^{\Psi}_{DD}}
{dM^2dp_T^2} =
\frac{\sigma_S}{16\pi B}\ .
\label{14c}
\eeq
\noi
Here $B\approx 5\ GeV^{-2}$ is the slope parameter of the 
differential diffractive cross section $hN \to X_{\Psi}N$.
Its value should be about a half of the slope of
elastic $NN$ scattering, since only the form factor of
the target nucleon contributes to the slope in diffraction to
large masses.

The parameter of the parameter $\sigma_S$
was fixed (using information
from deep-inelastic scattering) at $\sigma_S = 12\ mb$ 
in our analysis in the previous section. 
Since this value well describes the data on nuclear
suppression at high energy of $J/\Psi$,
we can safely use it on the same footing as the
data.

Thus, we arrive at the estimate,
\beq
\delta^{\Psi}_{DD}(M_m) 
\approx 0.12\ ,
\label{14d}
\eeq
\noi
which is close to the fraction of the diffractive
large rapidity gap events observed at HERA \cite{h1}.
This is not unexpected result, since, as we 
emphasized, there is much in common between deep-inelastic
scattering, Drell-Yan reaction and heavy flavor production,
especially when it concerns the role of soft interactions, which
dominate nuclear shadowing and diffraction.

One should not be confused with the fact that the left-hand 
sides of relations (\ref{14c}) and (\ref{14d}) depend on $M_m$,
while the right-hand sides are constant. The high-mass
tail $\propto 1/M_m^2$ of the diffractive $M^2$-distribution
leads to a logarithmic $q_c$-dependence of the nuclear formfactor,
which we neglected and eliminated the formfactor in
(\ref{14c}) and (\ref{14d}).

\section{Heavy ion collisions}

In this case either the fluctuations, or the formed charmonium
propagate through the both colliding nucleus. This may
sound puzzling, indeed, we considered above propagation
of the fluctuations of the projectile proton through the 
target nucleus. How such a fluctuation can propagate
through the projectile nucleus, which has the same velocity as the
projectile nucleon? This is a typical puzzle of the parton model.
It is impossible to prescribe a fluctuation either to the beam, or
to the target in a Lorentz-invariant way. It may be done only
in a concrete reference frame.

Thus, nuclear suppression in $AB$ collision is a product
of that in $pA$ and $pB$ interactions. Of course the 
kinematics should be treated properly, what is easier to do 
with Feynman variable $x_F = x_1-M^2_{\Psi}/x_1s$.
Then the nuclear suppression in AB collision reads \cite{hk95},
\beq
S_{AB}^{\Psi}(x_F) = S_{pA}^{\Psi}(x_F)\ 
S_{pB}^{\Psi}(-x_F) 
\label{15}
\eeq
Using expression (\ref{12}) - (\ref{13}) it easy to 
calculate the nuclear suppression factor $S_{AB}^{\Psi}(x_F)$
for $J/\Psi$, or applying eq.~(\ref{11}), for $\Psi'$.
It is interesting that due to inverse kinematics and
the strong $x_F$-dependence of $S_{pA}^{\Psi'}(x_F)$
it turns out that $S_{AB}^{\Psi'}(x_F) < 
S_{AB}^{\Psi}(x_F)$, in accordance with experimental observation.
However, the observed relative suppression of $\Psi'$
is even stronger than given by (\ref{15}). This is possible
due to final state interaction with the produced hadrons
(coomovers), which happens at long times when $\Psi'$ is 
formed and interacts with a large cross section.

The same coomovers are a plausible source of the additional
nuclear suppression of $J/\Psi$ production rate in $AB$-collisions,
which was found to be especially strong in lead-lead.
This is too complicate and uncertain problem to be discussed
in this note.

Summarising, we developed for the first time a phenomenological
approach to the problem of nuclear suppression of
charmonium production, based on the same dynamics, which
is responsible for nuclear shadowing in deep-inelastic scattering
and Drell-Yan reaction. The key point is an admixture of soft
mechanism of heavy flavour production, which scales in 
$M^2_{Q\bar Q}$, dominates at high $x_F$ and naturally
explains the observed strong nuclear suppression. 
Many other features observed experimentally are understood as well.
Our numerical estimates are in a good 
agreement with the fit to available data.

The $J/\Psi - N$ interaction cross section steeply grows with
energy due to increasing radius of the gluon cloud. 
As a result, the $J/\Psi$ production rate 
is expected to be very much  suppressed due to interaction with 
nucleons at very high energies. We expect $\sigma_{tot}^{\Psi N}
\approx 10\ mb$ at RHIC 
and $\sigma_{tot}^{\Psi N}\approx 26\ mb$ at LHC, 
if $J/\Psi$ is at rest in the c.m. 

{\bf Acknowledgements:} I benefited very much from discussions
and collaboration with J\"org H\"ufner during preparation of this talk. 
Some results presented here were obtained earlier in collaboration with
Omar Benhar and Andrzej Zieminski, and our numerous 
discussions were very helpful.

\end{document}